\documentclass[a4paper,twocolumn]{revtex4-1} 
\usepackage{amsmath}
\usepackage{graphicx}
\usepackage{hyperref}
\usepackage[utf8x]{inputenc}
\usepackage{hyperref}

\usepackage{fancyhdr}
\fancypagestyle{firstpage}{%
  \lhead{\footnotesize{\copyright 2019 IEEE.  Personal use of this material is permitted.  Permission from IEEE must be obtained for all other uses, in any current or future media, including reprinting/republishing this material for advertising or promotional purposes, creating new collective works, for resale or redistribution to servers or lists, or reuse of any copyrighted component of this work in other works.}}
  	\cfoot{}
	\lfoot{\footnotesize{This is the pre-peer reviewed version of the following article: V. V. Voronenkov, Y. S. Lelikov, A. S. Zubrilov, Y. G. Shreter and A. A. Leonidov, "Thick GaN Film Stress-induced Self-separation," \textit{2019 IEEE Conference of Russian Young Researchers in Electrical and Electronic Engineering (EIConRus)}, Saint Petersburg and Moscow, Russia, 2019, pp. 833-837. \href{https://doi.org/10.1109/EIConRus.2019.8657271}{DOI: 10.1109/EIConRus.2019.8657271}}}
  \rhead{}
}

\begin{document}


\title{Thick GaN film stress-induced self-separation }

\author{  Vladislav~Voronenkov$^{1}$ } \email{voronenkov@mail.ioffe.ru}
\author{   Andrey~Leonidov\textsuperscript{2}}
\author{   Yuri~Lelikov\textsuperscript{1}}
\author{   Andrey~Zubrilov\textsuperscript{1}}
\author{   Yuri~Shreter\textsuperscript{1}} \email{y.shreter@mail.ioffe.ru}

\affiliation{  
  \textsuperscript{1}\,Ioffe Physical-Technical Institute, St. Petersburg, Russia\\
  \textsuperscript{2}\,Peter the Great St. Petersburg Polytechnic University, St. Petersburg, Russia
 }

%

\begin{abstract}
Cracking of thick GaN films on sapphire substrates during the cooling down after the growth was studied.
The cracking was suppressed by increasing the film-to-substrate
thickness ratio and by using an intermediate carbon buffer layer, that reduced
the binding energy between the GaN film and the substrate. Wafer-scale self-separation of thick GaN films has been demonstrated.
\end{abstract}
\maketitle
\thispagestyle{firstpage}


\section{Introduction}

Gallium nitride substrate is the basis of modern high current density
devices: high-voltage diodes and transistors~\cite{ohta2015pndiode13GW,palacios2016vfet},
light emitting diodes~\cite{nakamura2015bulkLED}, vertical cavity
surface emitting lasers \cite{sony2016vcsel}, superluminescent light
emitting diodes~\cite{feltin2009SLED}, high electron mobility transistors
\cite{kuball2012HEMTthermo,kuball2012HEMT-nonarrenius,ammono2014HEMT}.
The main method for the production of GaN substrates today is the
hydride vapor phase epitaxy growth of bulk GaN film on a foreign
substrate, usually sapphire. An important step of the technological
process with this approach is the separation of the bulk GaN film
from the foreign substrate after growth, which is usually done by
removing the substrate with laser lift-off~\cite{kelly1999hvpe-liftoff,gogova2005llo},
or by stress-driven self-separation along the weakened interface between
the substrate and the film \cite{USUI-VAS-2003,HENNIG2008911-WSiN-ELOG}
or parallel to the interface~\cite{Natural-Stress-Moustakas2007,ASHRAF20102-FS-GaN-selfseparation,fujito2009-5mm,YAMANE20121}.
Cracking of the GaN film during the cooling down and during the self-separation
process is a significant problem, reducing the yield of the process.
Crack-free cooling of GaN films on a sapphire substrate was obtained
for films with a thickness of up to 300~$\mu$m, such films were
then separated from the substrate using the laser lift-off method~\cite{kelly1999hvpe-liftoff,gogova2005llo}.
Successful wafer-scale stress-induced self-separation has been reported
for GaN films with a thickness of $3\ldots5$~millimeters~\cite{fujito2009-5mm},
while cracking of GaN films with lower thickness was observed \cite{Natural-Stress-Moustakas2007}
unless special intermediate layers weakening the bonding between the
film and the substrate are used \cite{USUI-VAS-2003,HENNIG2008911-WSiN-ELOG}.
In this work, the processes of self-separation and cracking of
thick GaN films on sapphire substrates are investigated, and
the parameters for  the reproducible crack-free self-separation are determined.

\section{Experimental}

\subsection{Self-separation of GaN films on sapphire substrates}

A set of more than 100 GaN films  with a thickness from 100~$\mu m$
to 5000~$\mu m$ were grown on c-plane sapphire substrates with a thickness
of 430~$\mu m$ and a diameter of~52~$\mu m$~(2~inch). The two-stage
growth process was used to grow crack-free films with smooth surface
\cite{VVVvoronenkov2013two}. No special substrate surface treatment
to weaken the bonding between the substrate and the GaN film was employed. 

\begin{figure}
\includegraphics[width=1\columnwidth]{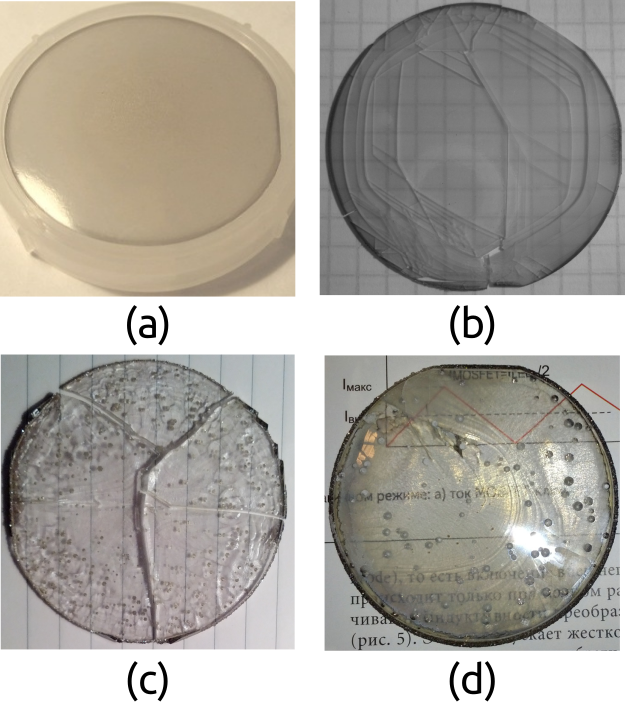}\caption{\label{fig:GaN-films}GaN films of various thickness $h_{GaN}$, grown
on sapphire substrates with a thickness of 430~$\mu m$ and a diameter
of 52~mm: a) $h_{GaN}$=200~$\mu$m, no cracks can be observed in the GaN
film and in the sapphire substrate, no signs of self-separation. b) $h_{GaN}$=400~$\mu$m, cracks in the GaN film and in the sapphire substrate were formed,
no self-separation. c)  $h_{GaN}$=2000~$\mu$m, GaN film self-separated
from the sapphire substrate over the entire area, several cracks were
formed in the GaN film dividing it into several pieces. d)  $h_{GaN}$=2800~$\mu$m,
GaN film self-separated from the sapphire substrate as a single piece.}

\end{figure}

The following typical failure modes were observed, depending on the
thickness of the GaN film:
\begin{itemize}
\item No self-separation occurred and no cracks were generated inside GaN
film and inside the sapphire substrate for GaN films thickness less
than $\sim$\,300~$\mu$m (fig. \ref{fig:GaN-films}a). 
\item Films with a thickness from $\sim$\,300~$\mu$m to $<$\,2500~$\mu$m
cracked into multiple pieces (fig. \ref{fig:GaN-films}b,c). Self-separation
of the GaN film from the substrate was observed in a plane located inside
the GaN film parallel to the interface. The separated area of the
film increased with increasing film thickness -- from the absence of separation for a 400-$\mu$m thick film to complete separation
for a 2000-$\mu$m thick film.
\item Films with a thickness higher than 2800~$\mu$m self-separated as a
single piece without cracking (fig. \ref{fig:GaN-films}d). The separation
plane was located inside the GaN film at a distance of 200-400~$\mu$m
from the interface.
\end{itemize}

\subsection{Self-separation of GaN film on a sapphire substrate  with a carbon buffer
layer}

\begin{figure}
\includegraphics[width=1\columnwidth]{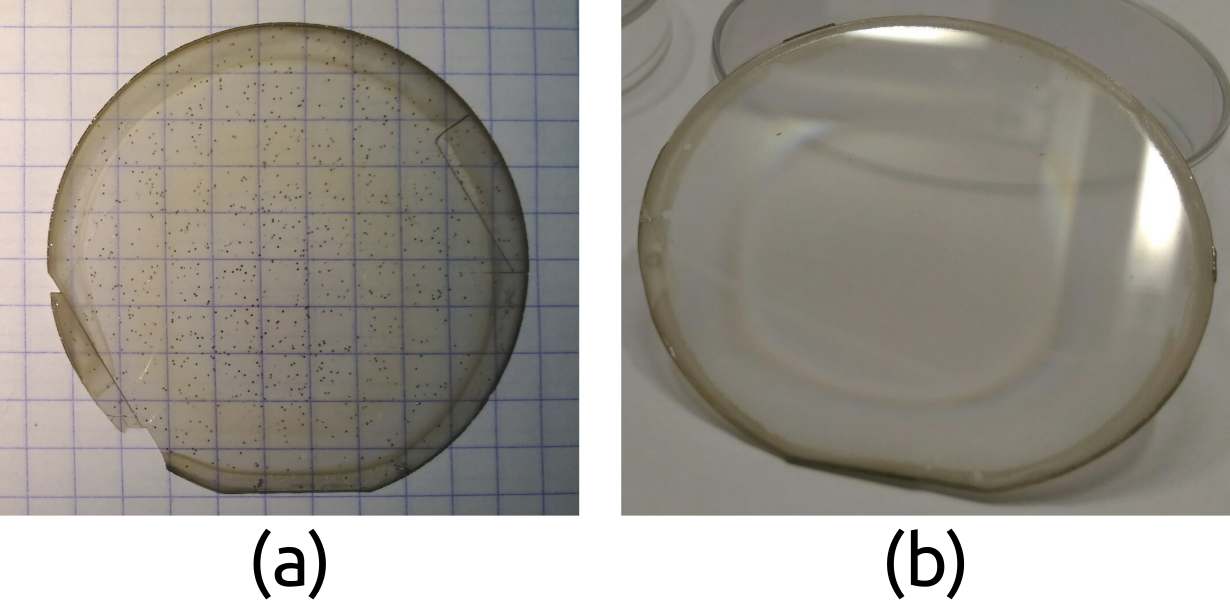}
\caption{\label{fig:carbon}
A 365-$\mu$m thick GaN film, grown on a 430~$\mu$m thick
sapphire substrate with a carbon buffer layer. a) The free-standing GaN film self-separated during
the cooling down as a single piece. b) The sapphire substrate  remained intact during the self-separation.
}
\end{figure}

GaN film with a thickness of 365~$\mu$m was grown epitaxially on
a sapphire substrate with a thickness of 430~$\mu$m using a carbon buffer
layer \cite{CARBON-Phil-2016}. The carbon buffer layer was deposited
by methane thermal decomposition process from a CH$_{4}$/H$_{2}$ mixture
\cite{BECKER1998-ch4-h2} at a deposition temperature of 1020~$^{\circ}$C,
a process pressure of 105~kPa and a CH\textsubscript{4} partial pressure
of 1.3~kPa. The GaN film self-separated as a single piece 
during cooling down after the growth (fig. \ref{fig:carbon}). Self-separation occurred along the interface of the GaN film and the substrate.

\section{Discussion}

\subsection{Crack types}

\begin{figure}
\includegraphics[width=1\columnwidth]{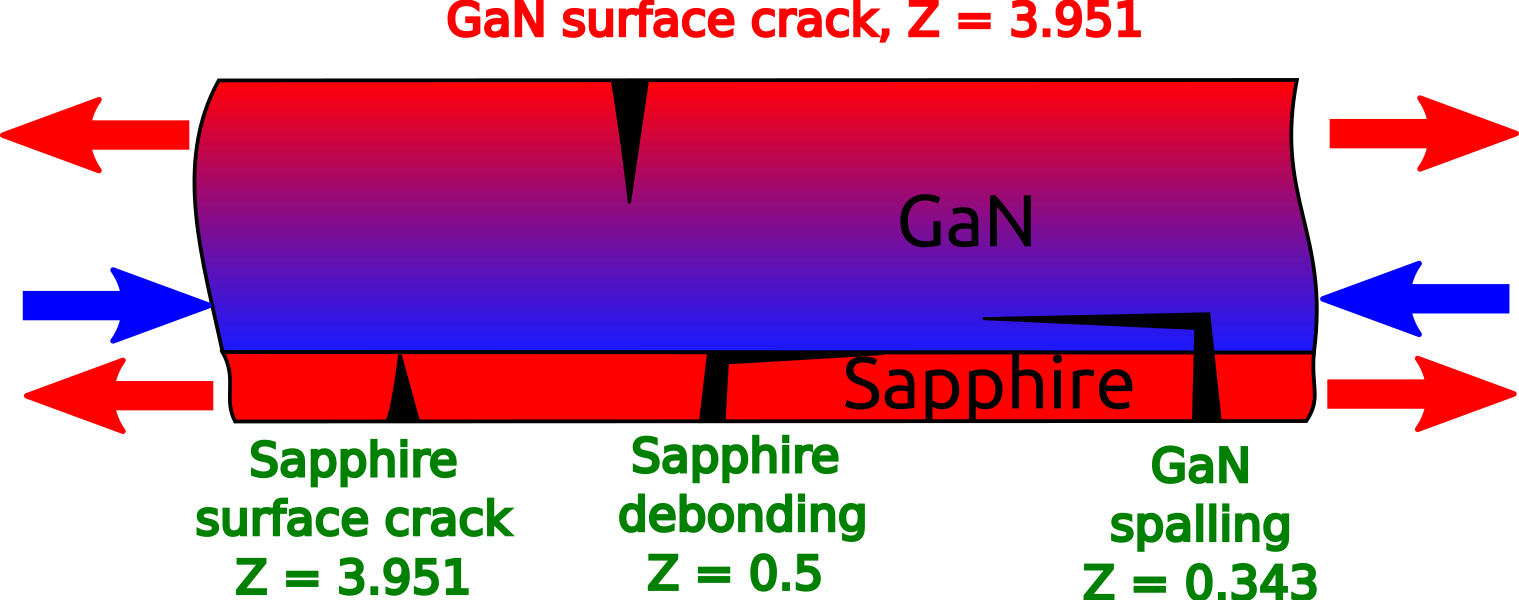}\caption{\label{fig:crack-patterns}
Schematic representation of the stress
distribution and the failure modes in a thick GaN film on a sapphire substrate ($h_{GaN}\gg h_{sapphire}$).
The stress in the GaN film is compressive near the interface and tensile
near the surface, the stress in the sapphire substrate is tensile across
the entire thickness. The values of a nondimensional driving force for
crack formation $Z$ are given for the case of semi-infinite GaN film
($h_{GaN}\gg h_{sapphire}$)~\cite{HUTCHINSON1991MMCLM}.}
\end{figure}

The following types of cracks were observed in thick GaN films, grown
on sapphire substrates, after cooling down: surface and channeling cracking of
sapphire and GaN, sapphire debonding along the interface and sapphire
debonding by  spalling (fig. \ref{fig:crack-patterns}). In case of spalling the crack propagates inside the GaN film parallel to the interface at a depth at which the in-plane shear stress intensity factor $K_{II}=0$ \cite{SUO1989spalling}. The steady-state spalling depth depends on the elastic mismatch between GaN and sapphire, and on the thicknesses of the GaN film and the sapphire substrate \cite{SUO1989spalling}. 

Sapphire debonding leads to self-separation of GaN film and is desirable or
at least permissible. Surface crack formation in the GaN film results
in cracking of the film into several parts during cooling down or
during further processing and is completely unacceptable.

To determine crack-free conditions for cooling and separation, the
stress distribution in the GaN film and the sapphire substrate and
the driving forces for crack formation were calculated.

\subsection{Stress distribution }

\begin{figure}
\includegraphics[width=1\columnwidth]{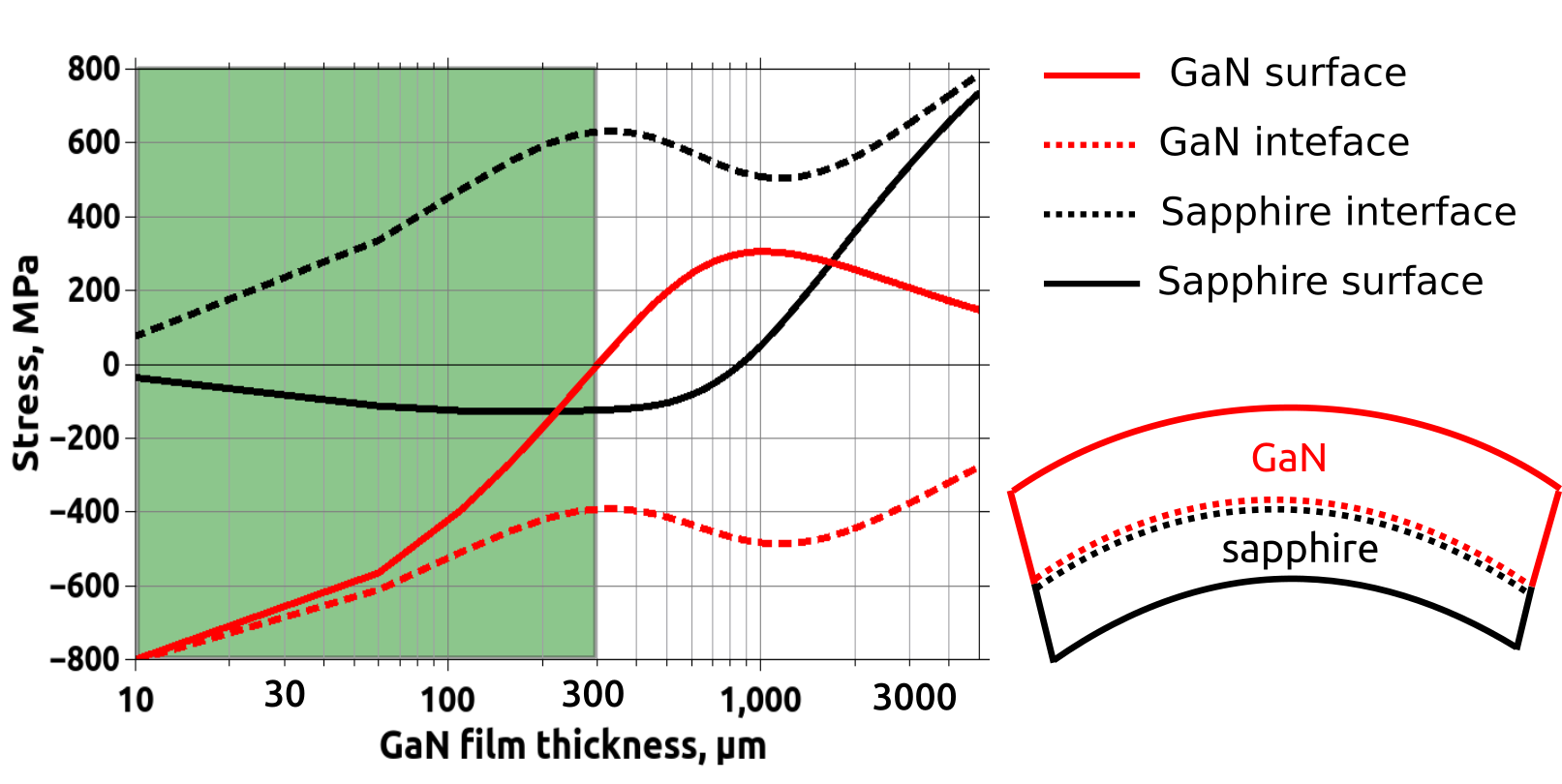}\caption{\label{fig:stress-distribution}
The stress in a GaN film on a sapphire
substrate with thickness of 430~$\mu$m and diameter of 52~mm as
a function of the GaN film thickness. When the GaN film thickness
is lower than 300~$\mu$m, the stress in the GaN film is compressive
and the formation of surface and channeling cracks  in the GaN film is
energetically unfavorable.}
\end{figure}

The deformation and distribution of stresses in a sapphire substrate
of radius $r$ and thickness $h_{sapphire}$, with a GaN film of thickness
$h_{GaN}$ were calculated by the minimization of the total strain energy,
assuming that the solution of the problem has radial symmetry. Elastic
properties of sapphire and GaN were assumed isotropic with the Young's
moduli of GaN and sapphire $E_{GaN}=343$~GPa and $E_{sapphire}=430$~GPa
and the Poisson's ratios $\nu_{GaN}=0.21$ and $\nu_{sapphire}=0.26$
\cite{deger1998gan-elastic,wagner2002properties,watchman1961sapphire-young,engel1960sapphire-poisson}.

The built-in strain $u_{0}$   was calculated taking into account the temperature dependence of the thermal expansion coefficients:
\begin{equation}
u_{0}=\int_{T_g}^{T_0}\left(\alpha_{GaN}(T)-\alpha_{sapphire}(T)\right) dT
\end{equation}
where $T_g$ is the growth temperature, $T_0$ is the room temperature $\alpha_{GaN}$ and $\alpha_{sapphire}$  are the thermal expansion coefficients of GaN and sapphire \cite{reeber2000lattice,krzhyzhanovsky1973thermophysical-oxydes}.

The components of the displacement vector in the interface plane were
taken as trial functions:

\begin{equation}
v_{r}\left(r\right)=k_{1}r+k_{3}r^{3}
\end{equation}

\begin{equation}
v_{z}(r)=k_{2}r^{2}+k_{4}r^{4}
\end{equation}
where $v_{r}$ -- radial displacement component, $v_{z}$ -- vertical
displacement component, $k_{1}\ldots k_{4}$ -- variation parameters.
Kirchhoff's hypothesis is used to approximate the displacement field:
\begin{equation}
u_{r}(r,z)=v_{r}-z\frac{\partial v_{r}}{\partial r}
\end{equation}

\begin{equation}
u_{z}(r,z)=v_{z}
\end{equation}

Nonlinear terms were taken into account when calculating the dependence
of the strain tensors on the displacement vector in cylindrical coordinates:

\begin{equation}
u_{rr}=\frac{\partial u_{r}}{\partial r}+\frac{1}{2}\left(\frac{\partial u_{r}}{\partial r}\right){}^{2}+\frac{1}{2}\left(\frac{\partial u_{z}}{\partial r}\right){}^{2}+u_{0}
\end{equation}

\begin{equation}
u_{\phi\phi}=\frac{u_{r}}{r}+\frac{1}{2}\left(\frac{u_{r}}{r}\right){}^{2}+u_{0},
\end{equation}

\begin{equation}
u_{zz}=\frac{-\nu}{(1-\nu)}\left(u_{rr}+u_{\phi\phi}\right),
\end{equation}

\begin{equation}
u_{r\phi}=\frac{1}{2}\left(\frac{\partial u_{r}}{\partial z}+\frac{\partial u_{z}}{\partial r}+\frac{\partial u_{r}}{\partial r}\frac{\partial u_{r}}{\partial z}+\frac{\partial u_{z}}{\partial r}\frac{\partial u_{z}}{\partial z}\right),
\end{equation}
 The stress tensor components and the specific elastic energy:

\begin{equation}
\sigma_{rr}=\frac{E}{1+\nu}\left(u_{rr}+\frac{\nu}{1-2\nu}\left(u_{rr}+u_{\phi\phi}+u_{zz}\right)\right),
\end{equation}

\begin{equation}
\sigma_{\phi\phi}=\frac{E}{1+\nu}\left(u_{\phi\phi}+\frac{\nu}{1-2\nu}\left(u_{rr}+u_{\phi\phi}+u_{zz}\right)\right),
\end{equation}

\begin{equation}
\sigma_{zz}=\frac{E}{1+\nu}\left(u_{zz}+\frac{\nu}{1-2\nu}\left(u_{rr}+u_{\phi\phi}+u_{zz}\right)\right),
\end{equation}

\begin{equation}
\sigma_{r\phi}=\frac{E}{1+\nu}u_{r\phi},
\end{equation}

\begin{multline}
U=\frac{E}{2\left(1+\nu\right)}((u_{rr}^{2}+u_{\phi\phi}^{2}+u_{zz}^{2}+2u_{r\phi}^{2})+\\
\frac{\nu}{1-2\nu}(u_{rr}+u_{\phi\phi}+u_{zz})^{2}).
\end{multline}

The total strain energy is obtained by integrating over volume:

\begin{equation}
F=\int_{0}^{R}\int_{-h_{sapphire}}^{h_{GaN}}UdV\label{eq:elasticity-total-energy}
\end{equation}

The minimum potential energy F is found from the solution of a system
of nonlinear equations:
\begin{equation}
\frac{\partial F}{\partial k_{i}}=0\label{eq:elasticity-nonlinear-equations}
\end{equation}

Analytical expressions for the system of nonlinear equations \ref{eq:elasticity-nonlinear-equations}
were obtained using the computer algebra system \cite{maxima}. The numerical
solution of the equation system \ref{eq:elasticity-nonlinear-equations}
was obtained using the Powell's method \cite{powell1964efficient}. The
results are shown in fig.~\ref{fig:stress-distribution}.

\subsection{Driving force for cracking }

\begin{figure}
\includegraphics[width=1\columnwidth]{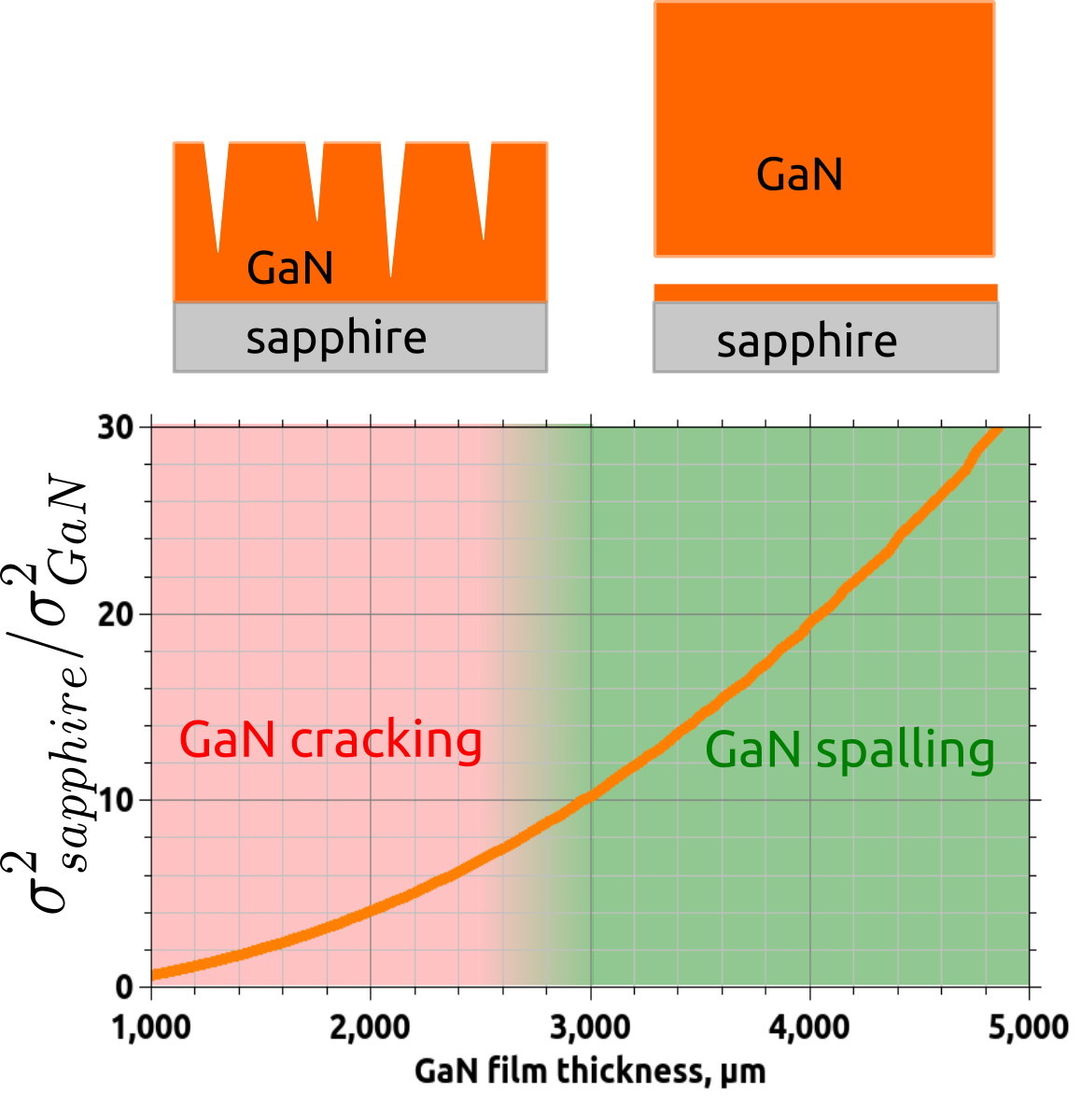}\caption{\label{fig:dr-force}The ratio of the substrate separation driving
force to the GaN cracking driving force: $\sigma{}_{sapphire}^{2}/\sigma{}_{GaN}^{2}$
on the thickness of the GaN film. The sapphire substrate thickness is 430~$\mu$m. 
Cracking of GaN films was observed at $\sigma{}_{sapphire}^{2}/\sigma{}_{GaN}^{2}\lesssim 6$, that corresponds to $h_{GaN}\lesssim$~2500~$\mu$m.
Reproducible separation without cracking was observed at $\sigma{}_{sapphire}^{2}/\sigma{}_{GaN}^{2}\gtrsim 8$, that corresponds to $h_{GaN}\gtrsim2800$~$\mu$m. }
\end{figure}

When the sapphire substrate thickness is $430\,\mu$m and the thickness
of the GaN film is less than $300\,\mu$m, the stress in the GaN film
and on the sapphire substrate surface is compressive (fig. \ref{fig:stress-distribution}), and the formation of surface and channeling cracks in the GaN film and in the sapphire substrate
is energetically unfavorable. 

When the GaN film thickness is higher than $300\,\mu$m, several competing
cracking processes shown in fig. \ref{fig:crack-patterns} take place.
What type of crack will arise first during the cooling process, depends
on the distribution of the stress in the substrate-film structure.
The driving force for cracking is the elastic strain energy
\begin{equation}
U=\frac{(1-\nu)}{E_{f}}\sigma^{2}
\end{equation}
 that is released during the crack formation. A crack of a certain type
can propagate if the energy release rate for this type of crack $G$
exceeds the surface energy of the formed crack $\Gamma$:
\begin{equation}
G=\frac{Z\sigma^{2}h}{E_{f}}>\Gamma
\end{equation}
where $Z$ is a dimensionless parameter depending on the crack geometry,
$\sigma$ -- stress in the film, $E_{f}$ -- Young's modulus of the
film, $h$ -- crack length \cite{HUTCHINSON1991MMCLM}. Sapphire substrate
cracking and separation by debonding and spalling is determined by
tensile stress in the substrate $\sigma_{sapphire}$, cracking of
GaN film is determined by tensile stress on the GaN film surface $\sigma_{GaN}$.
The ratio of energy release rates for spalling and for GaN cracking
$\frac{G_{spalling}}{G_{cracking}}$ is proportional to $\frac{\langle\sigma_{sapphire}^{2}\rangle}{\sigma_{GaN}^{2}}$.
With an increase in this ratio, it can be expected that the spalling
crack will be initiated and complete self-separation of the substrate
will occur before the stress on the surface of the GaN film exceeds
the fracture resistance. The dependence of the stress ratio $\frac{\langle\sigma_{sapphire}^{2}\rangle}{\sigma_{GaN}^{2}}$
on the thickness of the GaN film at a sapphire substrate thickness
of 430~$\mu$m is shown in  fig.~\ref{fig:dr-force}. The stress
ratio $\frac{\langle\sigma_{sapphire}^{2}\rangle}{\sigma_{GaN}^{2}}$
increases with increasing the GaN film thickness. Cracking of GaN films observed
at $h_{GaN}<$~2500~$\mu$m corresponds to $\frac{\langle\sigma_{sapphire}^{2}\rangle}{\sigma_{GaN}^{2}}\lesssim6$.
Reproducible separation without cracking observed at $h_{GaN}>$~2800~$\mu$m corresponds
to $\frac{\langle\sigma_{sapphire}^{2}\rangle}{\sigma_{GaN}^{2}}\gtrsim8$. 

Besides increasing the $\frac{\langle\sigma_{sapphire}^{2}\rangle}{\sigma_{GaN}^{2}}$
ratio by increasing the GaN film thickness, another way to promote
self-separation is to reduce the binding energy between the film and
the substrate. The surface energy of spalling crack in GaN is 6.7~J/m$^{2}$\cite{Li2015-GaN-surface-energy},
the use of the carbon buffer layer reduces the binding energy to 0.27~J/m$^{2}$\cite{Liu2012-graphite-binding},
which allowed to obtain wafer-scale self-separation of the GaN film
with thickness of 365~$\mu m$ (fig. \ref{fig:carbon}).

\section{Conclusion}

The thermal stress arising in a thick GaN film on a sapphire substrate
during the cooling down after growth is the driving force of the GaN
film cracking and the GaN self-separation. Crack-free freestanding
GaN film can be obtained in several ways:
\begin{itemize}
\item Growing GaN film with thickness $h_{GaN}<300$~$\mu$m. In this case,
the stresses on the free surfaces of the GaN film and the sapphire
substrate are compressive and the surface cracking process is energetically unfavorable.
The laser lift-off method can be used to separate the GaN film from the
sapphire substrate.
\item Growing GaN film with thickness $h_{GaN}>2800$~$\mu$m. In this
case the stress in the sapphire substrate is significantly higher
than the stress in the GaN film, and complete separation of the substrate
by spalling occurs before the stress in the GaN film exceeds the fracture
resistance.
\item Growing GaN films on an intermediate layer with a low binding energy. For
example, a  crack-free self-separation of a 365-$\mu$m thick GaN film
was demonstrated using a carbon buffer layer.
\end{itemize}
\bibliographystyle{IEEEtran}
\bibliography{lift-off-mechanics}

\begin{thebibliography}{10}
\providecommand{\url}[1]{#1}
\csname url@samestyle\endcsname
\providecommand{\newblock}{\relax}
\providecommand{\bibinfo}[2]{#2}
\providecommand{\BIBentrySTDinterwordspacing}{\spaceskip=0pt\relax}
\providecommand{\BIBentryALTinterwordstretchfactor}{4}
\providecommand{\BIBentryALTinterwordspacing}{\spaceskip=\fontdimen2\font plus
\BIBentryALTinterwordstretchfactor\fontdimen3\font minus
  \fontdimen4\font\relax}
\providecommand{\BIBforeignlanguage}[2]{{%
\expandafter\ifx\csname l@#1\endcsname\relax
\typeout{** WARNING: IEEEtran.bst: No hyphenation pattern has been}%
\typeout{** loaded for the language `#1'. Using the pattern for}%
\typeout{** the default language instead.}%
\else
\language=\csname l@#1\endcsname
\fi
#2}}
\providecommand{\BIBdecl}{\relax}
\BIBdecl

\bibitem{ohta2015pndiode13GW}
\BIBentryALTinterwordspacing
H.~Ohta, N.~Kaneda, F.~Horikiri, Y.~Narita, T.~Yoshida, T.~Mishima, and
  T.~Nakamura, ``{Vertical GaN p-n Junction Diodes With High Breakdown Voltages
  Over 4 kV},'' \emph{IEEE Electron Device Letters}, vol.~36, no.~11, pp.
  1180--1182, Nov 2015. [Online]. Available:
  \url{https://doi.org/10.1109/LED.2015.2478907}
\BIBentrySTDinterwordspacing

\bibitem{palacios2016vfet}
\BIBentryALTinterwordspacing
M.~Sun, Y.~Zhang, X.~Gao, and T.~Palacios, ``{High-Performance GaN Vertical Fin
  Power Transistors on Bulk GaN Substrates},'' \emph{IEEE Electron Device
  Letters}, vol.~38, no.~4, pp. 509--512, April 2017. [Online]. Available:
  \url{https://doi.org/10.1109/LED.2017.2670925}
\BIBentrySTDinterwordspacing

\bibitem{nakamura2015bulkLED}
\BIBentryALTinterwordspacing
C.~A. Hurni, A.~David, M.~J. Cich, R.~I. Aldaz, B.~Ellis, K.~Huang, A.~Tyagi,
  R.~A. DeLille, M.~D. Craven, F.~M. Steranka, and M.~R. Krames, ``{Bulk GaN
  flip-chip violet light-emitting diodes with optimized efficiency for
  high-power operation},'' \emph{Applied Physics Letters}, vol. 106, no.~3, p.
  031101, 2015. [Online]. Available: \url{https://doi.org/10.1063/1.4905873}
\BIBentrySTDinterwordspacing

\bibitem{sony2016vcsel}
\BIBentryALTinterwordspacing
T.~Hamaguchi, N.~Fuutagawa, S.~Izumi, M.~Murayama, and H.~Narui,
  ``{Milliwatt-class GaN-based blue vertical-cavity surface-emitting lasers
  fabricated by epitaxial lateral overgrowth},'' \emph{physica status solidi
  (a)}, vol. 213, no.~5, pp. 1170--1176, 2016. [Online]. Available:
  \url{https://doi.org/10.1002/pssa.201532759}
\BIBentrySTDinterwordspacing

\bibitem{feltin2009SLED}
\BIBentryALTinterwordspacing
E.~Feltin, A.~Castiglia, G.~Cosendey, L.~Sulmoni, J.-F. Carlin, N.~Grandjean,
  M.~Rossetti, J.~Dorsaz, V.~Laino, M.~Duelk, and C.~Velez, ``{Broadband blue
  superluminescent light-emitting diodes based on GaN},'' \emph{Applied Physics
  Letters}, vol.~95, no.~8, p. 081107, 2009. [Online]. Available:
  \url{https://doi.org/10.1063/1.3202786}
\BIBentrySTDinterwordspacing

\bibitem{kuball2012HEMTthermo}
\BIBentryALTinterwordspacing
N.~Killat, M.~Montes, J.~W. Pomeroy, T.~Paskova, K.~R. Evans, J.~Leach, X.~Li,
  U.~Ozgur, H.~Morkoc, K.~D. Chabak, A.~Crespo, J.~K. Gillespie, R.~Fitch,
  M.~Kossler, D.~E. Walker, M.~Trejo, G.~D. Via, J.~D. Blevins, and M.~Kuball,
  ``{Thermal Properties of AlGaN/GaN HFETs on Bulk GaN Substrates},''
  \emph{IEEE Electron Device Letters}, vol.~33, no.~3, pp. 366--368, March
  2012. [Online]. Available: \url{https://doi.org/10.1109/LED.2011.2179972}
\BIBentrySTDinterwordspacing

\bibitem{kuball2012HEMT-nonarrenius}
\BIBentryALTinterwordspacing
M.~Tapajna, N.~Killat, J.~Moereke, T.~Paskova, K.~R. Evans, J.~Leach, X.~Li,
  U.~Ozgur, H.~Morkoc, K.~D. Chabak, A.~Crespo, J.~K. Gillespie, R.~Fitch,
  M.~Kossler, D.~E. Walker, M.~Trejo, G.~D. Via, J.~D. Blevins, and M.~Kuball,
  ``{Non-Arrhenius Degradation of AlGaN/GaN HEMTs Grown on Bulk GaN
  Substrates},'' \emph{IEEE Electron Device Letters}, vol.~33, no.~8, pp.
  1126--1128, Aug 2012. [Online]. Available:
  \url{https://doi.org/10.1109/LED.2012.2199278}
\BIBentrySTDinterwordspacing

\bibitem{ammono2014HEMT}
\BIBentryALTinterwordspacing
P.~Kruszewski, P.~Prystawko, I.~Kasalynas, A.~Nowakowska-Siwinska, M.~Krysko,
  J.~Plesiewicz, J.~Smalc-Koziorowska, R.~Dwilinski, M.~Zajac, R.~Kucharski,
  and M.~Leszczynski, ``{AlGaN/GaN HEMT structures on ammono bulk GaN
  substrate},'' \emph{Semiconductor Science and Technology}, vol.~29, no.~7, p.
  075004, 2014. [Online]. Available:
  \url{https://doi.org/10.1088/0268-1242/29/7/075004}
\BIBentrySTDinterwordspacing

\bibitem{kelly1999hvpe-liftoff}
\BIBentryALTinterwordspacing
M.~K. Kelly, R.~P. Vaudo, V.~M. Phanse, L.~Gorgens, O.~Ambacher, and
  M.~Stutzmann, ``{Large Free-Standing GaN Substrates by Hydride Vapor Phase
  Epitaxy and Laser-Induced Liftoff},'' \emph{Japanese Journal of Applied
  Physics}, vol.~38, no.~3A, p. L217, 1999. [Online]. Available:
  \url{https://doi.org/10.1143/JJAP.38.L217}
\BIBentrySTDinterwordspacing

\bibitem{gogova2005llo}
\BIBentryALTinterwordspacing
D.~Gogova, C.~Hemmingsson, B.~Monemar, E.~Talik, M.~Kruczek, F.~Tuomisto, and
  K.~Saarinen, ``{Investigation of the structural and optical properties of
  free-standing GaN grown by HVPE},'' \emph{Journal of Physics D: Applied
  Physics}, vol.~38, no.~14, p. 2332, 2005. [Online]. Available:
  \url{https://doi.org/10.1088/0022-3727/38/14/007}
\BIBentrySTDinterwordspacing

\bibitem{USUI-VAS-2003}
\BIBentryALTinterwordspacing
Y.~Oshima, T.~Eri, M.~Shibata, H.~Sunakawa, K.~Kobayashi, T.~Ichihashi, and
  A.~Usui, ``{Preparation of Freestanding GaN Wafers by Hydride Vapor Phase
  Epitaxy with Void-Assisted Separation},'' \emph{Japanese Journal of Applied
  Physics}, vol.~42, no.~1A, p.~L1, 2003. [Online]. Available:
  \url{https://doi.org/10.1143/JJAP.42.L1}
\BIBentrySTDinterwordspacing

\bibitem{HENNIG2008911-WSiN-ELOG}
\BIBentryALTinterwordspacing
C.~Hennig, E.~Richter, M.~Weyers, and G.~Trankle, ``{Freestanding 2-in GaN
  layers using lateral overgrowth with HVPE},'' \emph{Journal of Crystal
  Growth}, vol. 310, no.~5, pp. 911 -- 915, 2008, proceedings of the E-MRS
  Conference, Symposium G. [Online]. Available:
  \url{https://doi.org/10.1016/j.jcrysgro.2007.11.102}
\BIBentrySTDinterwordspacing

\bibitem{Natural-Stress-Moustakas2007}
\BIBentryALTinterwordspacing
A.~D. Williams and T.~Moustakas, ``{Formation of large-area freestanding
  gallium nitride substrates by natural stress-induced separation of GaN and
  sapphire},'' \emph{Journal of Crystal Growth}, vol. 300, no.~1, pp. 37 -- 41,
  2007, first International Symposium on Growth of Nitrides. [Online].
  Available: \url{https://doi.org/10.1016/j.jcrysgro.2006.10.224}
\BIBentrySTDinterwordspacing

\bibitem{ASHRAF20102-FS-GaN-selfseparation}
\BIBentryALTinterwordspacing
H.~Ashraf, R.~Kudrawiec, J.~Weyher, J.~Serafinczuk, J.~Misiewicz, and
  P.~Hageman, ``{Properties and preparation of high quality, free-standing GaN
  substrates and study of spontaneous separation mechanism},'' \emph{Journal of
  Crystal Growth}, vol. 312, no.~16, pp. 2398 -- 2403, 2010. [Online].
  Available: \url{https://doi.org/10.1016/j.jcrysgro.2010.05.004}
\BIBentrySTDinterwordspacing

\bibitem{fujito2009-5mm}
\BIBentryALTinterwordspacing
K.~Fujito, S.~Kubo, H.~Nagaoka, T.~Mochizuki, H.~Namita, and S.~Nagao, ``{Bulk
  GaN crystals grown by HVPE },'' \emph{Journal of Crystal Growth}, vol. 311,
  no.~10, pp. 3011 -- 3014, 2009. [Online]. Available:
  \url{https://doi.org/10.1016/j.jcrysgro.2009.01.046}
\BIBentrySTDinterwordspacing

\bibitem{YAMANE20121}
\BIBentryALTinterwordspacing
K.~Yamane, M.~Ueno, H.~Furuya, N.~Okada, and K.~Tadatomo, ``{Successful natural
  stress-induced separation of hydride vapor phase epitaxy-grown GaN layers on
  sapphire substrates},'' \emph{Journal of Crystal Growth}, vol. 358, pp. 1 --
  4, 2012. [Online]. Available:
  \url{https://doi.org/10.1016/j.jcrysgro.2012.07.038}
\BIBentrySTDinterwordspacing

\bibitem{VVVvoronenkov2013two}
\BIBentryALTinterwordspacing
V.~Voronenkov, N.~Bochkareva, R.~Gorbunov, P.~Latyshev, Y.~Lelikov, Y.~Rebane,
  A.~Tsyuk, A.~Zubrilov, U.~Popp, M.~Strafela, and Y.~Shreter, ``{Two modes of
  HVPE growth of GaN and related macrodefects},'' \emph{Physica Status Solidi
  (c)}, vol.~10, no.~3, pp. 468--471, 2013. [Online]. Available:
  \url{https://doi.org/10.1002/pssc.201200701}
\BIBentrySTDinterwordspacing

\bibitem{CARBON-Phil-2016}
\BIBentryALTinterwordspacing
A.~S. Altakhov, R.~I. Gorbunov, L.~A. Kasharina, F.~E. Latyshev, V.~A. Tarala,
  and Y.~G. Shreter, ``{Amorphous carbon buffer layers for separating free
  gallium nitride films},'' \emph{Technical Physics Letters}, vol.~42, no.~11,
  pp. 1076--1078, Nov 2016. [Online]. Available:
  \url{https://doi.org/10.1134/S106378501611002X}
\BIBentrySTDinterwordspacing

\bibitem{BECKER1998-ch4-h2}
\BIBentryALTinterwordspacing
A.~Becker and K.~Huttinger, ``{Chemistry and kinetics of chemical vapor
  deposition of pyrocarbon - IV pyrocarbon deposition from methane in the low
  temperature regime},'' \emph{Carbon}, vol.~36, no.~3, pp. 213 -- 224, 1998.
  [Online]. Available: \url{https://doi.org/10.1016/S0008-6223(97)00177-2}
\BIBentrySTDinterwordspacing

\bibitem{HUTCHINSON1991MMCLM}
\BIBentryALTinterwordspacing
J.~Hutchinson and Z.~Suo, ``{Mixed Mode Cracking in Layered Materials},'' ser.
  Advances in Applied Mechanics, J.~W. Hutchinson and T.~Y. Wu, Eds.\hskip 1em
  plus 0.5em minus 0.4em\relax Elsevier, 1991, vol.~29, pp. 63 -- 191.
  [Online]. Available: \url{https://doi.org/10.1016/S0065-2156(08)70164-9}
\BIBentrySTDinterwordspacing

\bibitem{SUO1989spalling}
\BIBentryALTinterwordspacing
Z.~Suo and J.~W. Hutchinson, ``Steady-state cracking in brittle substrates
  beneath adherent films,'' \emph{International Journal of Solids and
  Structures}, vol.~25, no.~11, pp. 1337 -- 1353, 1989. [Online]. Available:
  \url{https://doi.org/10.1016/0020-7683(89)90096-6}
\BIBentrySTDinterwordspacing

\bibitem{deger1998gan-elastic}
\BIBentryALTinterwordspacing
C.~Deger, E.~Born, H.~Angerer, O.~Ambacher, M.~Stutzmann, J.~Hornsteiner,
  E.~Riha, and G.~Fischerauer, ``{Sound velocity of Al$_x$Ga$_{(1-x)}$N thin
  films obtained by surface acoustic-wave measurements},'' \emph{Applied
  Physics Letters}, vol.~72, no.~19, pp. 2400--2402, 1998. [Online]. Available:
  \url{https://doi.org/10.1063/1.121368}
\BIBentrySTDinterwordspacing

\bibitem{wagner2002properties}
\BIBentryALTinterwordspacing
J.~M. Wagner and F.~Bechstedt, ``{Properties of strained wurtzite GaN and AlN:
  Ab initio studies},'' \emph{Physical Review B}, vol.~66, no.~11, p. 115202,
  2002. [Online]. Available: \url{https://doi.org/10.1103/PhysRevB.66.115202}
\BIBentrySTDinterwordspacing

\bibitem{watchman1961sapphire-young}
\BIBentryALTinterwordspacing
J.~B. Wachtman, W.~E. Tefft, D.~G. Lam, and C.~S. Apstein, ``{Exponential
  Temperature Dependence of Young's Modulus for Several Oxides},'' \emph{Phys.
  Rev.}, vol. 122, pp. 1754--1759, Jun 1961. [Online]. Available:
  \url{https://doi.org/10.1103/PhysRev.122.1754}
\BIBentrySTDinterwordspacing

\bibitem{engel1960sapphire-poisson}
\BIBentryALTinterwordspacing
O.~G. Engel, ``{Resistance of White Sapphire and Hot Pressed Alumina to
  Collision With Liquid Drops},'' \emph{Journal of Research of the National
  Bureau of Standards}, vol.~64, no.~6, pp. 499--512, 1960. [Online].
  Available: \url{http://dx.doi.org/10.6028/jres.064A.049}
\BIBentrySTDinterwordspacing

\bibitem{reeber2000lattice}
\BIBentryALTinterwordspacing
R.~R. Reeber and K.~Wang, ``{Lattice parameters and thermal expansion of
  GaN},'' \emph{Journal of Materials Research}, vol.~15, no.~1, pp. 40--44,
  2000. [Online]. Available: \url{https://doi.org/10.1557/JMR.2000.0011}
\BIBentrySTDinterwordspacing

\bibitem{krzhyzhanovsky1973thermophysical-oxydes}
R.~Krzhyzhanovsky and Z.~Shtern, \emph{Thermophysical properties of
  non-metallic materials: oxides}.\hskip 1em plus 0.5em minus 0.4em\relax
  Energiya, 1973.

\bibitem{maxima}
\BIBentryALTinterwordspacing
Maxima. (2014) Maxima, a computer algebra system. version 5.34.1.
  http://maxima.sourceforge.net/. [Online]. Available:
  \url{http://maxima.sourceforge.net/}
\BIBentrySTDinterwordspacing

\bibitem{powell1964efficient}
\BIBentryALTinterwordspacing
M.~J.~D. Powell, ``{An efficient method for finding the minimum of a function
  of several variables without calculating derivatives},'' \emph{The Computer
  Journal}, vol.~7, no.~2, pp. 155--162, 1964. [Online]. Available:
  \url{https://doi.org/10.1093/comjnl/7.2.155}
\BIBentrySTDinterwordspacing

\bibitem{Li2015-GaN-surface-energy}
\BIBentryALTinterwordspacing
H.~Li, L.~Geelhaar, H.~Riechert, and C.~Draxl, ``{Computing Equilibrium Shapes
  of Wurtzite Crystals: The Example of GaN},'' \emph{Phys. Rev. Lett.}, vol.
  115, p. 085503, Aug 2015. [Online]. Available:
  \url{https://doi.org/10.1103/PhysRevLett.115.085503}
\BIBentrySTDinterwordspacing

\bibitem{Liu2012-graphite-binding}
\BIBentryALTinterwordspacing
Z.~Liu, J.~Z. Liu, Y.~Cheng, Z.~Li, L.~Wang, and Q.~Zheng, ``{Interlayer
  binding energy of graphite: A mesoscopic determination from deformation},''
  \emph{Phys. Rev. B}, vol.~85, p. 205418, May 2012. [Online]. Available:
  \url{https://doi.org/10.1103/PhysRevB.85.205418}
\BIBentrySTDinterwordspacing

\end{thebibliography}

\end{document}